# Characterisations of ohmic and Schottky contacts of a single ZnO nanowire

Bogdan Bercu[1], Wei Geng[2], Olivier Simonetti[1], Sergei Kostcheev[2], Corinne Sartel[3], Vincent Sallet[3], Gilles Lérondel[2], Michaël Molinari[1], Louis Giraudet[1], and Christophe Couteau[*,2,4]

[1] Laboratoire de Recherche en Nanosciences, Université de Reims Champagne-Ardenne, 51685 Reims, France.
[2] Laboratoire de Nanotechnologie et d'Instrumentation Optique, Institut Charles Delaunay, CNRS UMR 6279, Université de Technologie de Troyes, 10000, Troyes, France.
[3] Groupe d'étude de la matière condensée (GEMAC), CNRS – Université de Versailles St Quentin, 78035 Versailles Cedex, France.
[4] CINTRA UMI 3288, CNRS-Thales-NTU, Research Techno Plaza, 50 Nanyang Drive, 637553 Singapore.

**Abstract**: *Current–voltage and Kelvin Probe Force Microscopy (KPFM) measurements were performed on single ZnO nanowires. Measurements are shown to be strongly correlated with the contact behavior, either ohmic or Schottky. The ZnO nanowires were obtained by metallo-organic chemical vapor deposition (MOCVD) and contacted using electronic-beam lithography. Depending on the contact geometry, good quality ohmic contacts (linear I-V behavior) or non-linear (diode-like) Schottky contacts were obtained. Current-voltage and KPFM measurements on both types of contacted ZnO nanowires were performed in order to investigate their behavior. A clear correlation could be established between the I-V curve, the electrical potential profile along the device and the nanowire geometry. Some arguments supporting this behavior are given based on a depleted region extension.. This work will help to better understand the electrical behavior of ohmic contacts on single ZnO nanowires, for future applications in nanoscale field effect-transistors and nano-photodetectors.*

**Introduction**

As the low dimensionality and the high surface to volume ratio have a major impact in the overall photodetection performance, there has been an increased interest in the recent years for the use of semiconductor nanowires as photodetectors. A wide variety of semiconductor materials have been used, such as Si [1], GaN [2], GaAs [3], ZnO [4-7], Ge [8,9], ZnTe [10], CdS [11], SnO$_2$ [12], InN [13] and W$_{18}$O$_{49}$ [14]. Some of the reported results prove that single nanowires can actually exhibit strong photoconduction gain. Among them, ZnO nanowires have been of particular interest as photoconduction gains as high as $10^{10}$ [7] in the UV region were reported. Such a high gain means that the photon-electron conversion can reach several orders of magnitude in a single nanowire which can be sufficient to detect single photons or at least to detect very low intensity light. ZnO nanowires seem thus ideal for applications requiring high spatial resolution (sub-micron resolution) and low optical power detection such as single molecule applications for biology or for quantum optics experiments.

Nevertheless, several issues come into play when it comes to making a precise measurement of the photoconductive gain of a single nanowire. An important issue is the good quality of the injecting contacts on the nanowire and the reproducibility of its characteristics. Current-voltage (I-V) measurements of biased nanowires [15] and local surface potential mapping using KPFM [16-26] are generally used in order to characterize the contacted nanowires. In this article, we present the electrical characterization results of ZnO nanowires obtained by MOCVD and contacted by Ti/Au pads using e-beam lithography [27].

**Experimental**

ZnO nanowires were grown by MOCVD on c-sapphire substrates at 577K under 85 Torr. The densely packed vertical ZnO nanowires were then sonicated in an ethanol solution in order to remove them from the initial substrate. Single nanowires are obtained by drop-casting the resulting solution on a silicon substrate with a top $SiO_2$ insulating layer of 400 nm (estimated by ellipsometry measurements). Electrical contacts on single nanowires were obtained using a lift-off technique based on e-beam lithography on PMMA resist. Using a mechanical indenter, an alignment grid is drawn on the surface to perform the lithography in a single step. More details on the fabrication process can be found in [27]. The photoluminescence spectra of single nanowires are first measured using a microphotoluminescence setup in order to select good quality ZnO nanowires. The selected ones were the ones presenting high photoluminescence intensity and good waveguiding effect along the nanowire, thus insuring a fairly good structural NW. The indenter marks are then used as alignment marks for the e-beam lithography which will define the contact pads for the nanostructures (figure 1b). We finally evaporated 30 nm of titanium and 300 nm of gold on the substrate followed by the lift-off procedure.

The contacted nanowires were characterized using current-voltage and surface potential measurements. The electrical characterizations were performed using an Agilent 5270B analyzer and were done in the dark to avoid any spurious effect due to the light environment. The surface potential measurements on biased ZnO nanowires have been performed using the Kelvin Probe Force Microscopy technique (KPFM) [28, 29]. For that, a Dimension 3100 atomic force microscope (AFM) was used in the lift mode, in a controlled atmosphere environment. The measurement method consists in two consecutive probe scans for each raster scan line: the first measures the topography of the sample and the second the local potential using the Kelvin probe technique [30].

**Results and discussion**

ZnO nanowires with diameters between 120 to 250 nm and length between 11 to 16 µm were measured (such as shown in Figure 1).

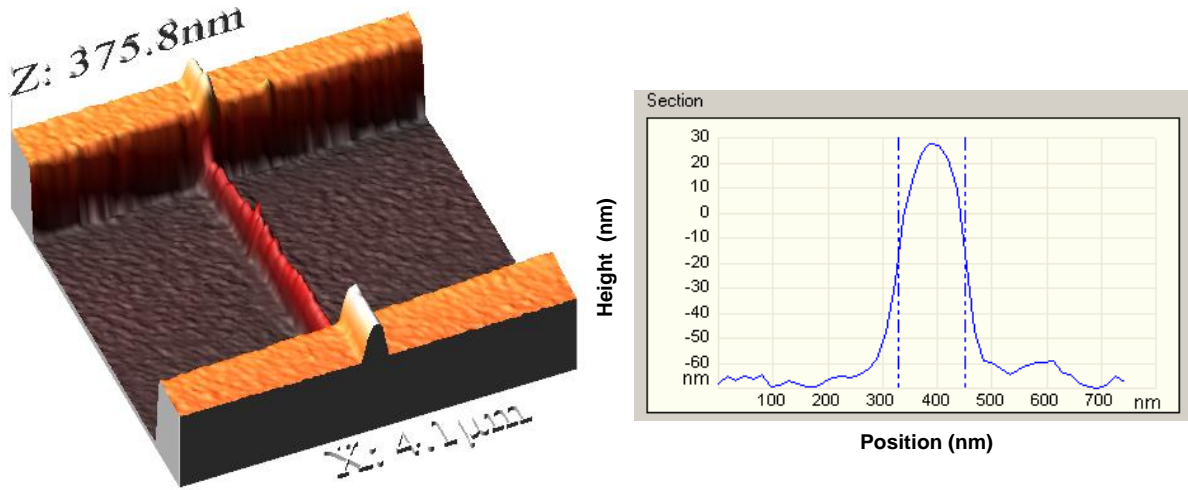

**Figure 1.** AFM 3D image of a contacted ZnO nanowire (left) with a corresponding cross section (right) showing an average nanowire width of 130nm.

As depicted Figure 2 most nanowires exhibit linear I-V behavior, showing that the technological process provides good ohmic contacts.

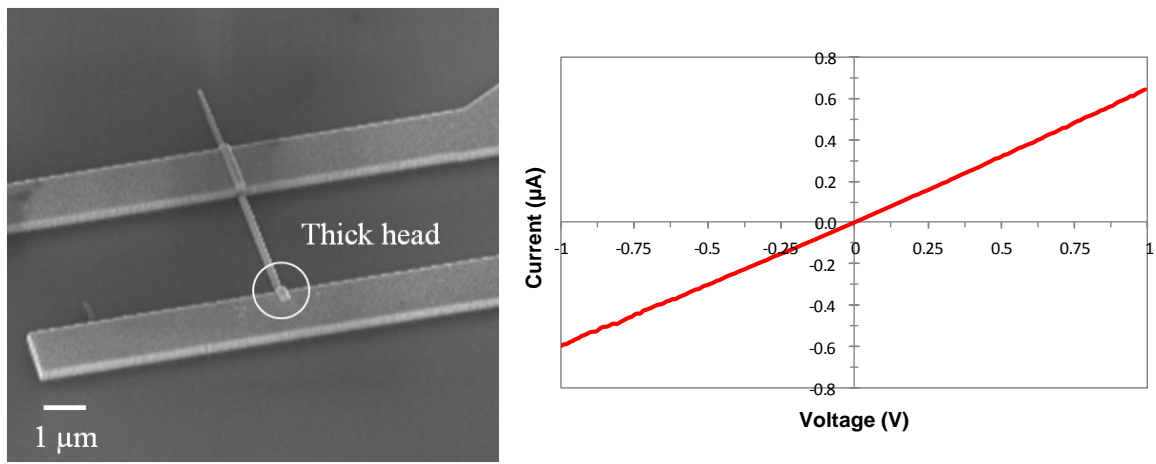

**Figure 2.** SEM image (left) and current-voltage measurement (right) of a ZnO nanowire showing linear (ohmic) behavior.

Surface potential maps of the linear nanowire were obtained using the KPFM technique for different bias voltages. Figure 3 presents the KPFM measurements for a 1V bias applied in a forward (Fig. 3a) and reverse (Fig. 3b) direction. As can be observed in the longitudinal potential cross-section graph (Fig. 4), the potential distribution along the ZnO nanowire with linear contacts is identical regardless of the bias voltage application direction, as expected from ohmic contacts. Also a linear potential profile is clearly observed along the whole nanowire between the contacts. From the I(V) measurements, neglecting the ohmic contact series resistances contributions, the nanowire resistivities are estimated to be between 0.23 and 2.4 Ω.cm (with resistances ranging from 0.25 to 2.4 MΩ), for an electrode spacing of 4 µm and an average nanowire diameter of 200 nm [27]. Upon applying a gate voltage (between + 10 V and -10 V, not shown here), we found that the ZnO NWs were n-doped, which is in good agreement with standard residual doping in ZnO [31].

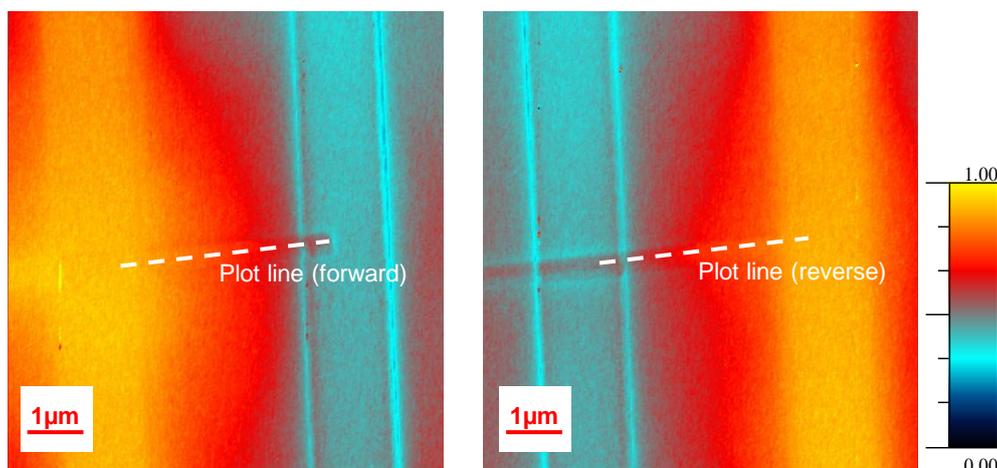

**Figure 3.** Potential measurements on a contacted ZnO nanowire with linear I-V behavior for a forward (left) and reverse (right) bias (U=1V). A symmetric behavior is observed.

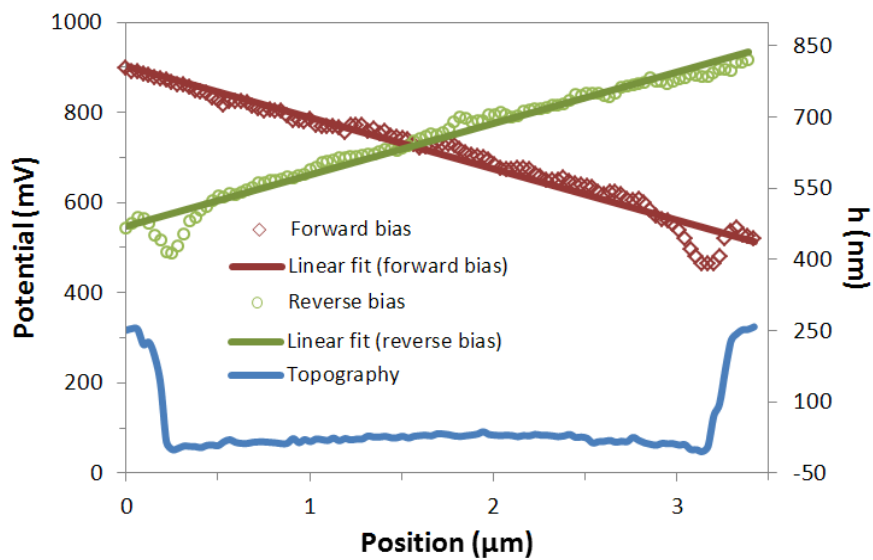

**Figure 4.** Measured topography and electrical potential variation along a ZnO nanowire with ohmic contacts.

A diode-like I-V behavior was also found for a small fraction of the contacted nanowires (Figure 5) due to the presence of one Schottky-behaving Au-ZnO contact.

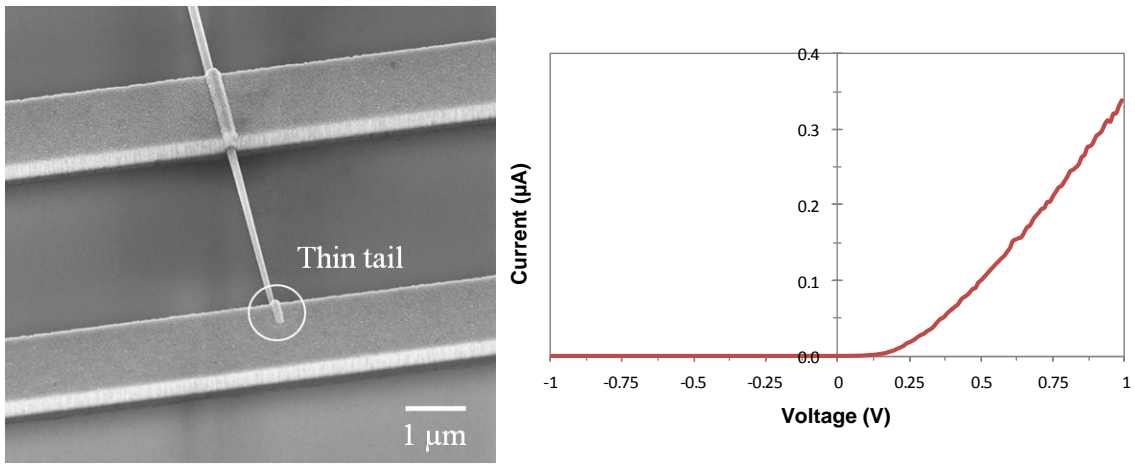

**Figure 5.** SEM image (left) and I-V measurement (right) on a non-ohmic ZnO nanowire.

The surface potential of such Schottky-behaving contacts was also investigated using KPFM measurements. Figure 6 presents the generated surface potential distributions for forward (Fig 6.a) and reverse (Fig 6.b). As can be seen from the longitudinal cross-section along the nanowire (Fig. 7), the potential profile is linear for the forward bias (similar to Figure 4) and highly non-linear for reverse bias. Fig. 8 represents a schematic band diagram of a Schottky contact. From this picture, one can see that when a forward bias is applied, the voltage drop in the depletion region of the Schottky contact is very limited: the applied bias is then quasi-linearly distributed over the whole nanowire length. On the contrary, under reverse bias, the depletion region at the Schottky contact extends into the ZnO volume, and most of the applied voltage is lost at this region. Consequently, the measurement shows an almost constant potential distribution along the nanowire, but a sharp voltage drop at the ZnO nanowire Schottky contact vicinity. This situation is illustrated Fig. 8 (right).

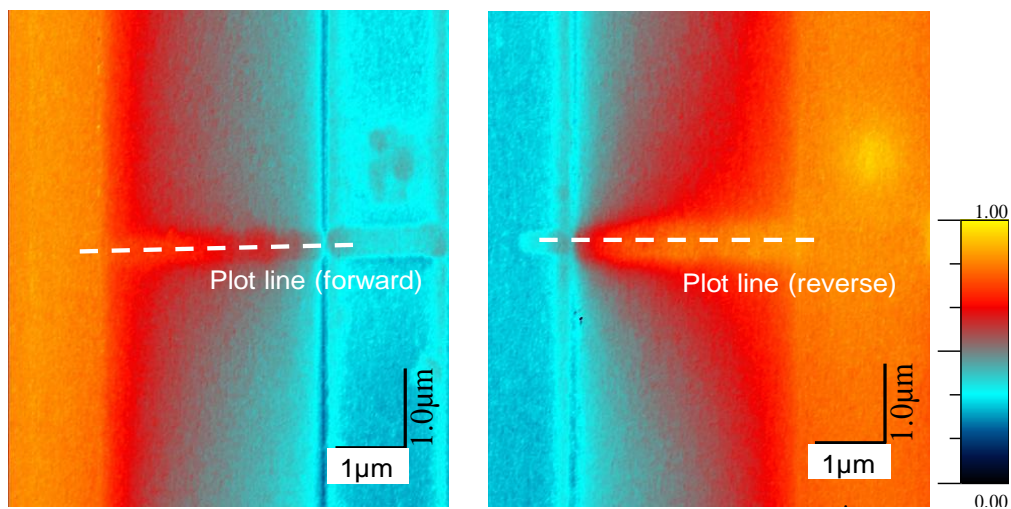

**Figure 6.** Potential measurements on a contacted ZnO nanowire with diode-like I-V behavior for a forward (left) and reverse (right) bias (U=1V).

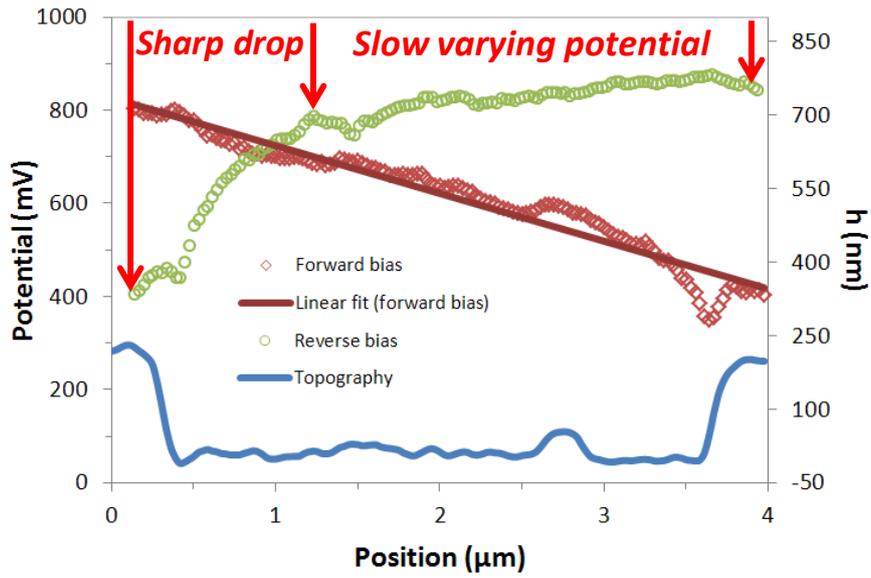

**Figure 7.** Measured topography and electrical potential variation along a ZnO nanowire with a Schottky contact.

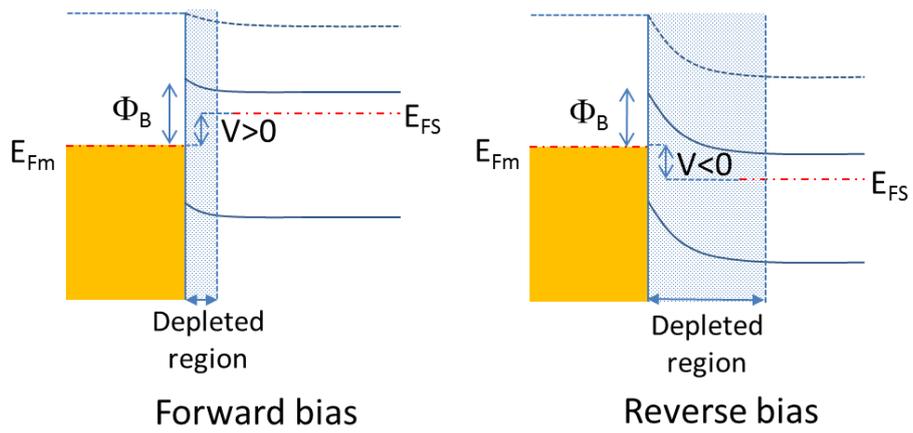

**Figure 8.** Schematic band diagram of a Schottky contact under forward (left) and reverse (right) bias. $\Phi_B$ is the barrier height at the metal-semiconductor contact, $E_{Fm}$ and $E_{FS}$ are respectively the Fermi level in the metal and in the semiconductor. The grey areas represent the extensions of the depleted region at the metal / semiconductor interface.

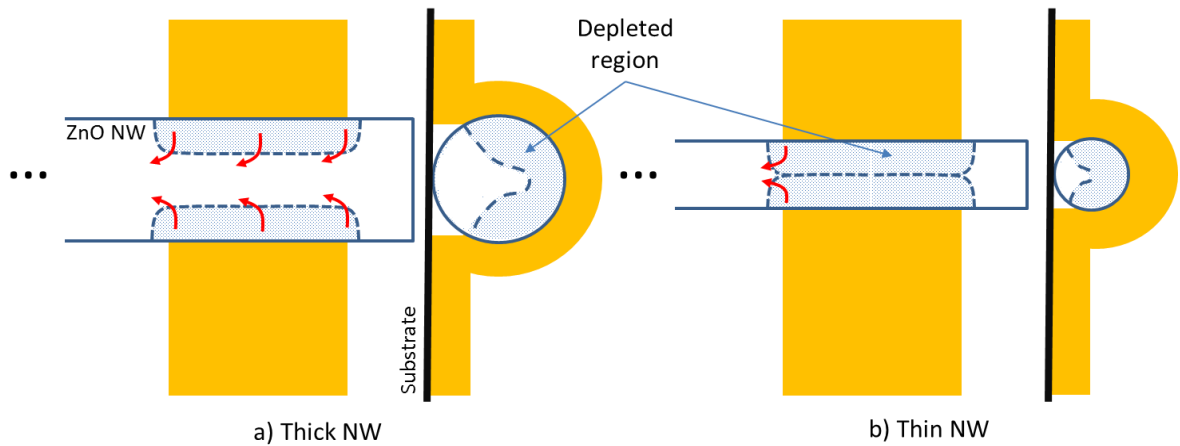

**Figure 9.** Schematic top and side views of thick (a) and thin (b) ZnO nanowires and their gold contacts under reverse bias, illustrating the effect of the depleted region extension. The schematic arrows indicate current flow under the partially or fully depleted contacts.

A possible explanation for the ohmic or Schottky contact behavior follows the observation of the nanowire shape. Due to the growth conditions, all the nanowires are slightly needle shaped. From quantitative head and tail diameter study using an electron beam microscope, it turned out that the head to tail diameter ratio of the nanowires are above 70% for all the ohmic behaving contacts, and decreases to between 50 and 70% for the Schottky behaving contacts. As a matter of fact, the Schottky-behaving contact in all the diode-like nanowires, was always located on the narrow side of the wire. Some fabrication issues cannot be discarded due to the very thin needle end of these nanowires, such as lithography, metal deposition, or very limited contact surface between the metal and the ZnO.

Another phenomenon may emphasize the non-linear behavior in narrow wires, if one considers a residual depletion region being present at the contact/ZnO interface, as illustrated Fig.8 and Fig.9, due to an energy barrier. When the nanowire diameter reduces below twice the depletion region extension, the nanowire may become fully depleted under the contact and highly isolating (Fig.9 b): the I(V) behavior in this situation may become highly non linear, as observed in reverse and low forward bias. On the contrary, for large diameter nanowires (Fig.9 a) the depletion region never extends over the whole diameter and the current can flow from the entire metal contact surface into the nanowire: the I(V) behaves linearly. This explanation is reinforced by the observation of some intermediate situations, where very leaky Schottky (or poor ohmic) contacts were measured.

**Conclusions**

Current-voltage characterization and Kelvin probe potential mapping were performed on single nanowire devices made of ZnO, showing a good ohmic I(V) behavior correlated to a linear distribution of the electric potential along the nanowire. Some Schottky-behaving nanowire devices were also identified and characterized. In this case, KPFM potential measurements revealed an abrupt potential drop at the Schottky contact vicinity, as expected from the presence of a depletion region in such contacts. This behavior could be correlated to the needle shape of the nanowires, and in particular a diode-like I(V) curve is observed for nanowires exhibiting one very narrow end. Non-linear I(V) behavior observed in narrow nanowires is thought to be enhanced due to a depleted zone extending over the entire ZnO volume under the contact.